\documentclass{article}

\usepackage{PRIMEarxiv}

\usepackage[utf8]{inputenc} 
\usepackage[T1]{fontenc}    
\usepackage{hyperref}       
\usepackage{url}            
\usepackage{booktabs}       
\usepackage{amsfonts}       
\usepackage{nicefrac}       
\usepackage{microtype}      
\usepackage{lipsum}
\usepackage{fancyhdr}       
\usepackage{graphicx}       
\usepackage{siunitx}
\usepackage[numbers]{natbib}
\usepackage{todonotes}
\usepackage{subcaption}
\usepackage{tikz,xcolor,hyperref}

\makeatletter
\def\@email#1#2{%
 \endgroup
 \patchcmd{\titleblock@produce}
  {\frontmatter@RRAPformat}
  {\frontmatter@RRAPformat{\produce@RRAP{*#1\href{mailto:#2}{#2}}}\frontmatter@RRAPformat}
  {}{}
}%
\makeatother
\graphicspath{{media/}}     

\DeclareRobustCommand{\orcidicon}{%
    \begin{tikzpicture}
    \draw[lime, fill=lime] (0,0) 
    circle [radius=0.16] 
    node[white] {{\fontfamily{qag}\selectfont \tiny ID}};    \draw[white, fill=white] (-0.0625,0.095) 
    circle [radius=0.007];    \end{tikzpicture}
    \hspace{-2mm}}
\foreach \x in {A, ..., Z}{%
    \expandafter\xdef\csname orcid\x\endcsname{\noexpand\href{https://orcid.org/\csname orcidauthor\x\endcsname}{\noexpand\orcidicon}}
    }
\title{Martini 3 application for the design of bistable nanomachines
\thanks{\textit{\underline{Citation}}: 
\textbf{Authors. Title. Pages.... DOI:000000/11111.}} 
}

\author{
  Alexander D. Muratov\orcidA, Vladik A. Avetisov\orcidB \\
  Semenov Federal Research Center for Chemical Physics, Russian Academy of Sciences, 119991 Moscow, Russia\\
  Design Center for Molecular Machines, Moscow, Russia \\
  \texttt{alexander.muratov@chph.ras.ru~(A.D.M.); avetisov@chph.ras.ru~(V.A.A.);} \\
}

\begin{document}
\maketitle

\begin{abstract}
During our previous modeling using all-atom molecular dynamics, we have identified several foldamers whose nanoscale behavior resembles that of classic bistable machines, namely the Euler archs and Duffing oscillators. However, time limitations of the all-atom molecular dynamics prevent us from performing a full-scale investigation of long-time behavior and prompt us to develop a coarse-grained model. In this work, we summarize our recent research on developing such models using the most widely available method called Martini.
\end{abstract}


\section{Introduction}
\label{sec:I}
Recent developments in the construction of molecular nanomachines require candidates for compounds that can act as two-state systems, since they are useful in various areas of nanomachine design, from switches and sensors to energy harvesters\cite{mi6081046,C5SC02317C,zhang_molecular_2018,thibado_fluctuation-induced_2020}. Our previous all-atom modeling\cite{avetisov2019oligomeric,avetisov2021short} has identified poly(N-isopropylacrylamide) (PNIPA) and pyridine-furan (PF) foldamers to exhibit behavior reminiscent of bistable rods and springs such as an Euler arch\cite{arnold_catastrophe_1984,poston_catastrophe_1996} and Duffing oscillator\cite{duffing1918erzwungene,korsch_duffing_2008}, respectively. These short oligomeric systems demonstrate clear bistable behavior with thermally activated spontaneous vibrations (SV) and stochastic resonance (SR). 

\begin{figure}
    \centering
    \includegraphics[width=\textwidth]{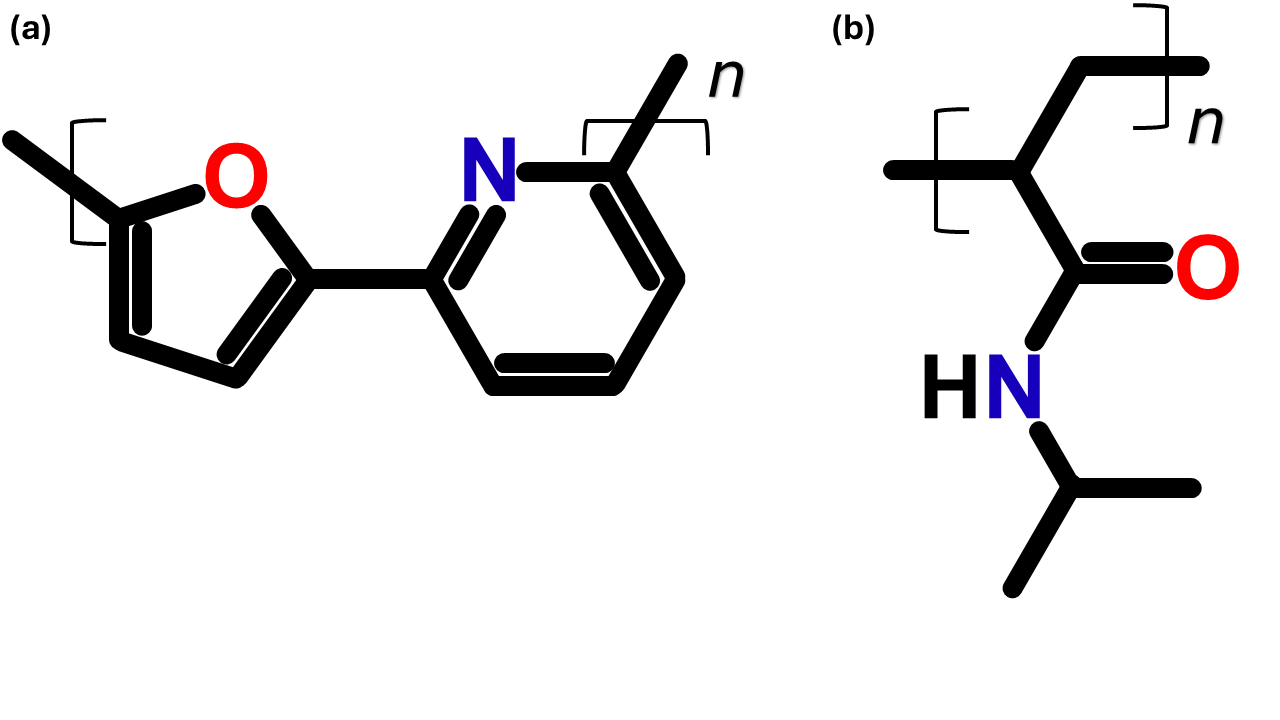}
    \caption{Chemical structure of (a) cis-pyridine-furan monomer unit (b) poly(N-isopropylacrylamide) monomer unit.}
    \label{fig:fig1}
\end{figure}
PNIPA (see Figure \ref{fig:fig1}(a)for more details) is a thermosensitive compound that experiences a reversible "coil-to-globule" transition at its lower critical solvation temperature (LCST) around $\SI{305}{\kelvin}$\cite{kamath_thermodynamic_2013,pasparakis_lcst_2020}. Its thermoresponsive nature is usually explained by the fact that amide groups below LCST tend to form hydrogen bonds with water, and other amide groups within polymer above LCST. Our previous atomistic modeling has revealed that thermoresponsive molecules may demonstrate behavior resembling that of an Euler arch\cite{avetisov2019oligomeric}, well-known mechanical example of a one-dimensional (1D) bistable system. Basically being an elastic rod, Euler arch remains straight when squeezed in the longitudinal direction, but as soon as the shrinkage force exceeds a critical value, it bifurcates into two symmetrical curved states. Furthermore, an application of a lateral force may trigger jump-like transitions between these two states. Within our model, we observed similar switch-like random transitions driven by a lateral force and spontaneous vibrations between two states under the bifurcation point, and these vibrations were activated by thermal noise. An additional weak oscillating force would transform spontaneous vibrations into stochastic resonance: regular, but still noise-induced transitions\cite{avetisov2019oligomeric,markina2020detection}.

 Another type of compounds that may exhibit spontaneous vibrations and stochastic resonance according to the results of all-atom molecular dynamics, are nanosprings such as oligomeric pyridine-furan (oligo-PF). PF is a conductive polymer composed of 5- and 6-member heterocyclic rings (consult Figure \ref{fig:fig1}(b) for more details)\cite{ALANJONES19968707,jones1997extended,sahu2015}. $\pi$-$\pi$ interactions between aromatic rings may lead to nonlinear elasticity of the oligmeric PF spring, which is in line with the calculations of density functional theory calculations\cite{sahu2015,tsuzuki_accuracy_2020}. Indeed, we have once again observed thermally activated spontaneous vibrations and stochastic resonance while modeling a cis-configuration of the pyridine-furan (PF) spring with five monomer units (oligo-PF-$5$) under stretching\cite{avetisov2021short,astakhov_spontaneous_2024}. Our findings support the idea that the dynamics of such springs may correspond to those of Duffing oscillator - a system whose bistable dynamics are characterized by damped oscillations of nonlinearly elastic springs. Moreover, through classical molecular dynamics modeling, we have shown that coupling of several oligo-PF springs by an oligomeric bridge or graphene plate leads to spontaneous synchronization of their vibrations\cite{markina_spontaneous_2023,frolkina_collective_2025}.

Both systems are of interest for the design of molecular machines, and using molecular dynamics simulations we have highlighted their possible use as, for example, sensing elements\cite{markina2020detection}. Relatively large constructions capable of collective bistability have also been considered\cite{markina_spontaneous_2023,frolkina_collective_2025}. However, although we observe bistable behavior in such constructions, we cannot help but notice that the increase in the number of bistable nanosprings leads to a longer average lifetime in each state of a collectively bistable system, which is also confirmed by our theoretical findings\cite{markina_spontaneous_2023,frolkina_collective_2025,neiman_synchronizationlike_1994}. Thus, the lifetimes and length scales of larger systems may be unavailable by means of atomistic molecular dynamics. In such a case, the methods of coarse-grained (CG) molecular dynamics can be of use, of which the so-called Martini (usually deciphered as "MARrink Toolkit INItiative") stands out as it offers comparable to atomistic molecular dynamics precision combined with significantly improved performance\cite{marrink_martini_2007}, especially in the recent version coined Martini $3.0$\cite{souza_martini_2021,alessandri_martini_2022}.

Lately, we have already parametrized both PNIPA and pyridine-furan within the Martini framework\cite{muratov_modeling_2021,muratov_coarse-graining_2024}. However, the Martini model for PNIPA is now slightly outdated as we used an early beta version of Martini $3.0$ and it requires reparameterization; as for oligo-PF-$5$, the findings of our atomistic modeling\cite{astakhov_spontaneous_2024,frolkina_collective_2025} indicate that it manifests better bistable behavior in the hydrophobic solvent, tetrahydrofuran (THF). Thus, the objectives of the current article are the following: to update the Martini model for PNIPA and to confirm that oligo-PF-$5$ preserves its bistable behavior in THF.

\section{Materials and Methods}
\label{sec:MM}
\begin{figure}
    \centering
    \includegraphics[width=\textwidth]{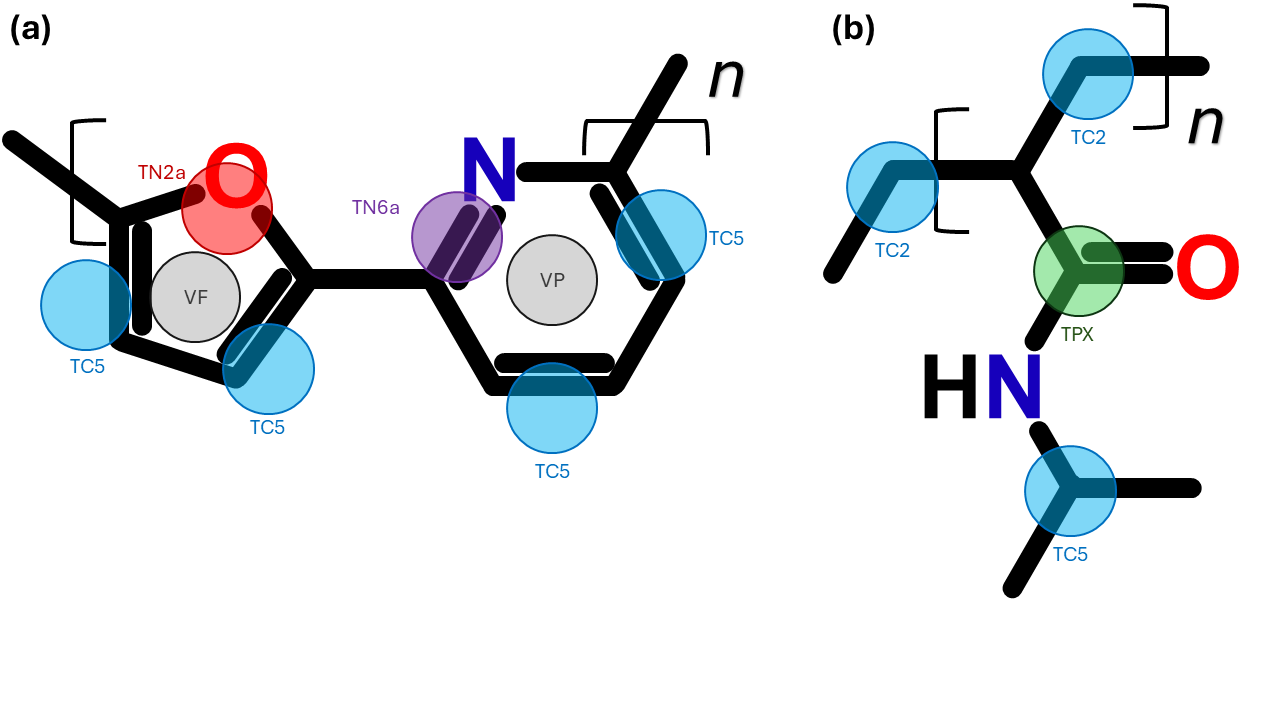}
    \caption{Mapping scheme of (a) cis-pyridine-furan monomer unit (b) poly(N-isopropylacrylamide) monomer unit.}
    \label{fig:fig2}
\end{figure}
\subsection{Bonded and non-bonded parameterization}\label{sec:MM1}
The procedures for parameterization of both non-bonded and bonded interactions were described in our previous works; here, we briefly summarize the data represented there\cite{muratov_modeling_2021,muratov_coarse-graining_2024}. Martini applies a "building block" strategy for the description of chemical compounds, utilizing beads of discrete types (charged (Q), polar (P), nonpolar (N), apolar (C), divalent (D), halogen-containing (X), and water (W)) and with discrete sizes (tiny(T), small(S), and regular(R)). Most bead types are divided into subtypes; therefore, the Lennard-Jones interaction between beads of different subtypes is also discretized; there are $22$ possible interaction levels\cite{souza_martini_2021}. To describe oligo-PF-$5$, we use preparameterized rigid structures for both furan and pyridine; then, we adopt the "Divide and Conquer" strategy for their connection\cite{alessandri_martini_2022,muratov_coarse-graining_2024} (see Figure \ref{fig:fig2} (a) for details of mapping). For PNIPA, we first have to calculate non-bonded parameters based on partition free energies, which we report in Section \ref{sec:R2}; for bonded parameters, we refer to the previously published data; in short, we adopt the "A-mapping" strategy for the description of the polymer chain\cite{muratov_modeling_2021,rossi_coarse-graining_2011}(see Figure \ref{fig:fig2}(b) for mapping details).

\subsection{Computational details}\label{sec:MM2}
For modeling of both oligo-PF-$5$ and PNIPA we use Gromacs $2023$\cite{berendsen_gromacs_1995,lindahl_gromacs_2001,van_der_spoel_gromacs_2005,hess_gromacs_2008,pronk_gromacs_2013,pall_tackling_2015,abr2015,abraham_gromacs_2023} maintaining the number of particles, volume, and temperature constant (NVT ensemble). For oligo-PF-$5$ we set the temperature to $\SI{298}{\kelvin}$; for PNIPA we set the temperature to $\SI{280}{}$/$\SI{330}{\kelvin}$ to reproduce its behavior below/above LCST. In each case, the temperature is kept constant during simulation by a velocity rescale thermostat\cite{bussi2007canonical} with $\SI{1.0}{\pico\second}$ coupling time. 

For non-bonded parameterization of PNIPA we perform thermodynamic integration (TI) to calculate partitioning free energies using DLPOLY Classic (version $1.9$) following the procedure described in our previous work\cite{muratov_modeling_2021}.

In our investigation of the oligo-PF-$5$’s bistable behavior, we conducted experiments where one end of the spring was secured while a tensile force was exerted on the opposite end along its longitudinal axis. We used the end-to-end distance, represented as $R_{e}$, as a collective coordinate to characterize the long-term dynamical properties of the spring. The bistable nature of the oligo-PF-5 spring was confirmed by the presence of two distinct and consistently observed states, characterized by end-to-end separations of approximately $R_{e}\sim \SI{0.4}{\nano\metre}$ and $R_{e}\sim \SI{0.75}{\nano\metre}$. These configurations correspond to the squeezed and the stress-strain states of the spring, respectively.

\section{Results}
\label{sec:R}
\subsection{Spontaneous vibrations and stochastic resonance of oligo-PF-5 in hydrophobic solvent}\label{sec:R1}
\begin{figure}
  \centering
  \includegraphics[width=\textwidth]{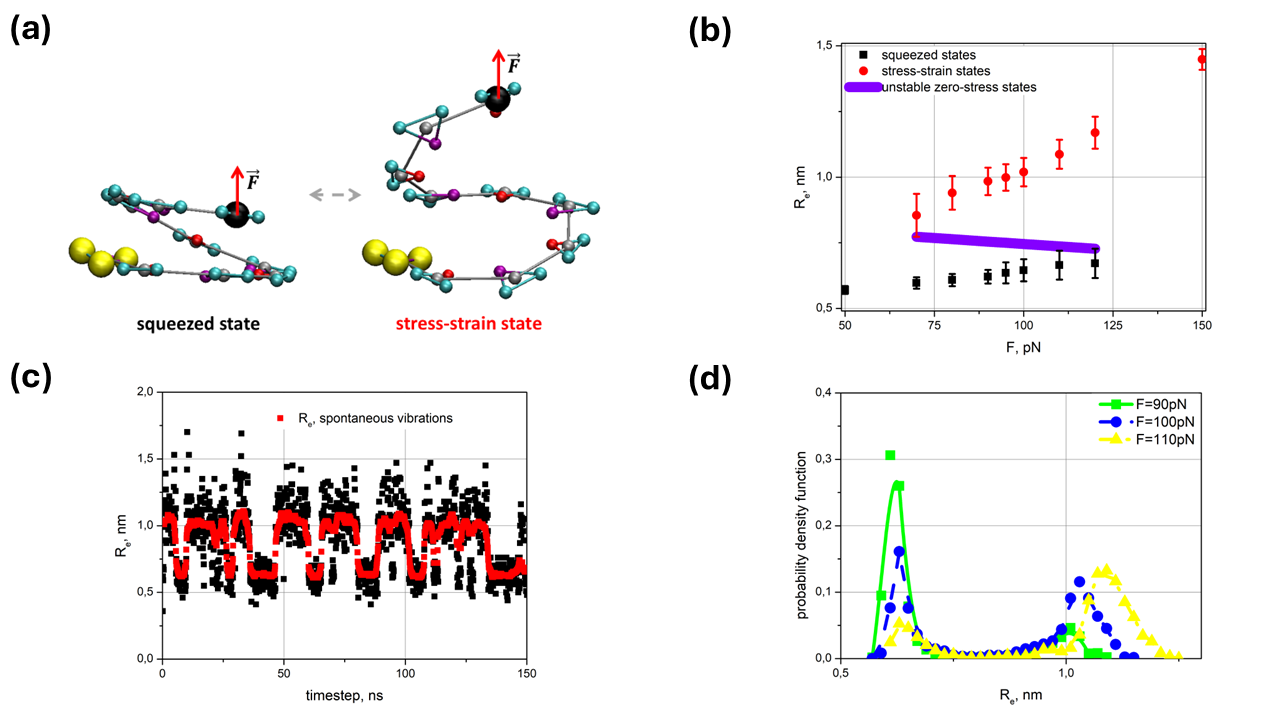}
  \caption{(a) Coarse-grained model of the oligo-PF-$5$ system with a pulling force. The squeezed and the stress–strain states of the spring are shown on the left and right, respectively. The yellow spheres at the lower end of the spring indicate the fixation of the pyridine ring by rigid harmonic force. The pulling force, $F$, is applied to the top end of the spring. (b) The state diagram shows a linear elasticity of oligo-PF-$5$ spring up to  $F\approx \SI{70}{\pico\newton}$ and bistability of the spring in the region from $F\approx\SI{70}{}- \SI{120}{\pico\newton}$; (c) Spontaneous vibrations of the oligo-PF-$5$ spring at  $F\approx \SI{100}{\pico\newton}$; (d) Evolution of the probability density for the squeezed and stress–strain states when pulling force surpasses the critical value.}
  \label{fig:fig3}
\end{figure}

To investigate the bistable response of the oligo-PF-$5$ spring to tensile forces in THF, we used a methodology similar to that used for water\cite{avetisov2021short,muratov_coarse-graining_2024}. After achieving thermal equilibrium at $\SI{298}{\kelvin}$ with one end of the spring stationary, we introduced a pulling force $\vec{F}$ directed along the longitudinal axis of the spring. Under low tension conditions, the spring maintained its initial configuration. However, upon exceeding a critical force threshold of about $F_c = \SI{70}{\pico\newton}$, the spring exhibited spontaneous transitions between two distinct states: a compressed form and a stretched conformation. The difference in mean end-to-end distances between these states amounted to roughly $\SI{0.35}{\nano\metre}$ in agreement with both $ab$ $initio$ calculations\cite{sahu2015} and atomistic modeling\cite{astakhov_spontaneous_2024}. Visual representations of these two states at the coarse-grained level are presented in Figure \ref{fig:fig3}(a).

The transition dynamics between the squeezed and stress-strain states of the oligo-PF-$5$ spring as the applied force surpasses the critical value $F_{c}$ are illustrated in Figures \ref{fig:fig2}(b,d). Within the force range of $F$ from $90$ to $\SI{110}{\pico\newton}$, both the squeezed and stress-strain states are visited with nearly identical probabilities. Figure \ref{fig:fig2}(c) demonstrates a representative time series of the end-to-end distance $R_{e}(t)$ for the oligo-PF-$5$ spring within the symmetric bistability region, with average lifetimes of approximately $\tau = \SI{8.5}{\nano\second}$ for both states.

\begin{figure}
  \centering
  \includegraphics[width=\textwidth]{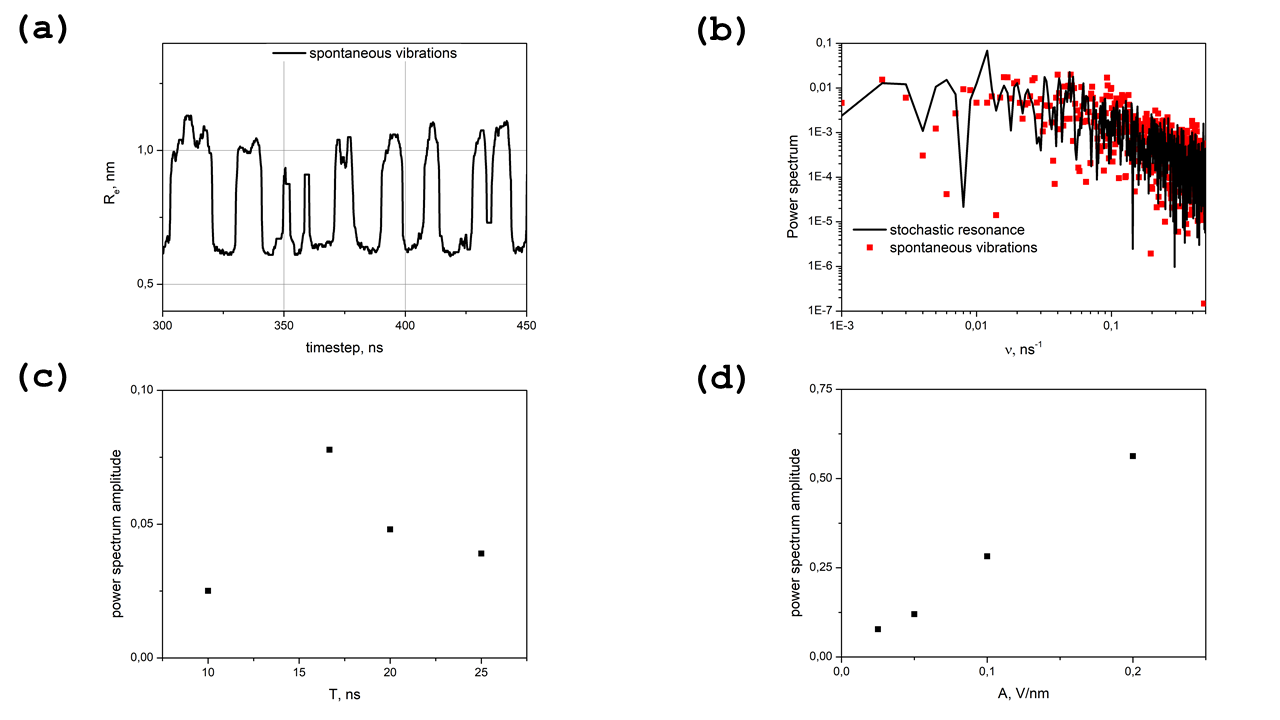}
  \caption{Stochastic resonance of the oligo-PF-$5$ induced by an oscillating field $E = E_{0} \cos (2 \pi \nu t)=E_{0} \cos (\nicefrac{2 \pi t}{T})$: (a) The dynamic trajectory at $F = \SI{100}{\pico\newton}$, $T = \SI{16.67}{\nano\second}$, and $E_{0} = \SI{0.025}{\volt\per\nano\metre}$; (b) Power spectra of spontaneous vibrations (red curve) and stochastic resonance (black curve); (c) The dependence of the main resonance peak amplitude on the period $T$ of oscillating field ($E_{0}=\SI{0.025}{\volt\per\nano\metre}$); (d) The dependence of the main resonance peak amplitude on $E_{0}$ ($T_{0}=\SI{16.67}{\nano\second}$).}
  \label{fig:fig4}
\end{figure}
To investigate the stochastic resonance behavior of the oligo-PF-5 spring, we introduced a weak oscillating electric field described by $E = E_0 \cos(2\pi \nu t)$, acting on a unit charge located at the pulled end of the spring. A characteristic temporal evolution of the oligo-PF-$5$ spring under stochastic resonance conditions is presented in Figure \ref{fig:fig4}(a). The corresponding power spectral densities for both spontaneous oscillation and stochastic resonance regimes are illustrated in Figure \ref{fig:fig4}(b). The principal resonance peak was observed at the frequency $\nu=(2\tau)^{-1}$, where the period of the external oscillatory field matches twice the mean residence time of a state in the spontaneous oscillation regime, according to stochastic resonance theory\cite{gammaitoni_stochastic_1998,wellens_stochastic_2004}.We conducted experiments with different oscillatory fields to determine the conditions for stochastic resonance, and the findings are summarized in Figures \ref{fig:fig4}(c,d).

\subsection{Partitioning free energies of PNIPA}\label{sec:R2}
\begin{table*}
\small
\caption{$\log{K_{OW}}$ for different types of beads representing AM (horizontal) and IP (vertical) groups of PNIPA for Martini $3.0$\label{tab:tab1} }
\begin{tabular*}{\textwidth}{@{\extracolsep{\fill}}c|ccc|ccc|ccc|ccc}
\hline
&\multicolumn{3}{c|}{$\SI{280}{K}$}&\multicolumn{3}{c|}{$\SI{300}{K}$}&\multicolumn{3}{c|}{$\SI{310}{K}$}&\multicolumn{3}{c}{$\SI{330}{K}$}\\
&\multicolumn{3}{c|}{$\log{K_{OW}}=0.48$}&\multicolumn{3}{c|}{$\log{K_{OW}}=0.8$}&\multicolumn{3}{c|}{$\log{K_{OW}}=1.4$}&\multicolumn{3}{c}{$\log{K_{OW}}=1.5$}\\
&TP$3$&TP$4$&TP$5$&TP$3$&TP$4$&TP$5$&TP$3$&TP$4$&TP$5$&TP$3$&TP$4$&TP$5$\\
\hline
TC$4$&$---$&$---$&$0.29$&$---$&$---$&$0.72$&$---$&$---$&$0.7 $&$---$&$---$&$1.2 $\\
TC$5$&$1.14$&$0.25$&$0.06$&$1.69$&$0.74$&$0.29$&$1.67$&$0.91$&$0.5 $&$1.84$&$1.08$&$0.61$\\
\hline
\end{tabular*}
\end{table*}
To parameterize PNIPA within Martini $3.0$, we have to recalculate partitioning free energies between water and octanol-$1$. Technically speaking, partitioning free energies are not the only target parameters for Martini $3.0$; however, our previous results were in good agreement with atomistic data, and since the main difference between the legacy beta version of Martini and the current one is the LJ parameters, we focus on adjusting them.

As in our earlier work, for reference, we use partitioning free energies in the CHARMM atomistic force field\cite{kamath_thermodynamic_2013}. To start with, we use a TC$2$ bead for the representation of the PNIPA backbone, and that remains unchanged during the selection of parameters. This choice may seem arbitrary, but it corresponds to the general Martini $3.0$ parameterization rules and is also in line with the data available on other similar polymers\cite{souza_martini_2021,rossi_coarse-graining_2011,banerjee_coarse-grained_2018}. Hence, we have to determine the exact CG beads for the amide and isopropyl groups. We start with a TP$5$ bead and TC$5$, respectively, and vary the bead subtypes, calculating the free energy of partitioning for each combination by thermodynamic integration. The results of these calculations are given in Table \ref{tab:tab1}; these results suggest several possible options. Both TC$2$-TP$4$-TC$5$ and TC$2$-TP$5$-TC$4$ representations are possible considering the calculation error of $\SI{0.4}{\kilo\joule\per\mole}$, while TC$2$-TP$3$-TC$5$ fits the atomistic data at temperatures above LCST. Moreover, different representations may be suitable at different temperatures, as in our previous work\cite{muratov_modeling_2021}
\subsection{Choice of the representation at different temperatures}\label{sec:R3}

\begin{figure}
  \centering
  \includegraphics[width=\textwidth]{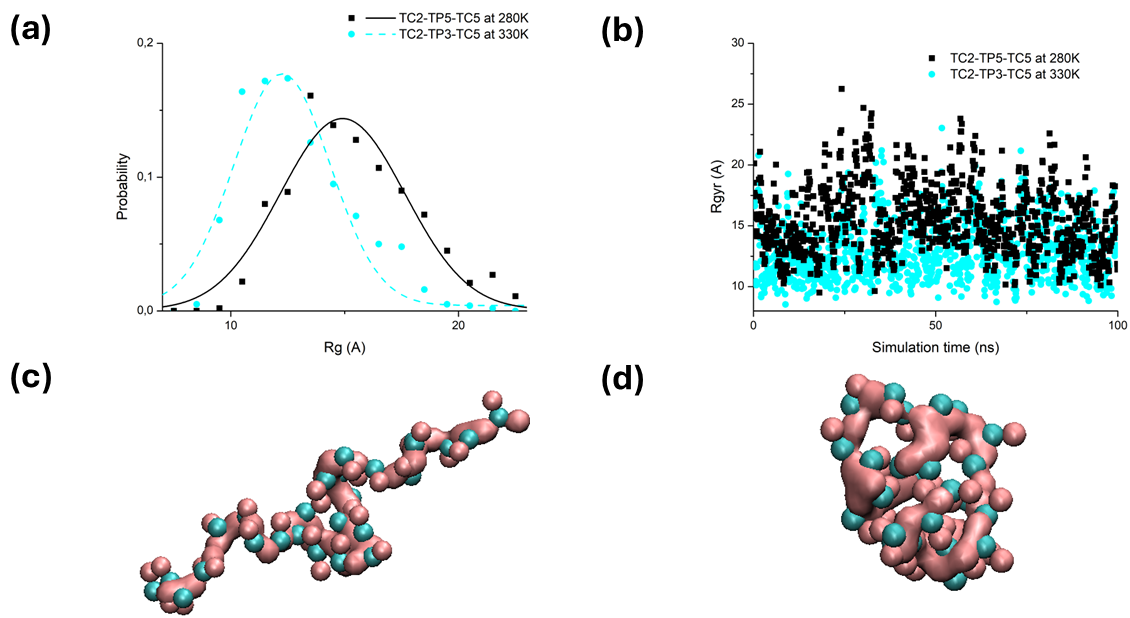}
 \caption{The representation of isotactic PNIPA of $30$ monomer units: the probability distribution of the radius of gyration at (a) $\SI{280}{\kelvin}$ and $\SI{330}{\kelvin}$; (b) the radii of gyration vs time for isotactic PNIPA at $\SI{280}{\kelvin}$ and $\SI{330}{\kelvin}$ with $N=30$ monomer units; snapshot of PNIPA at (c) $\SI{280}{\kelvin}$; (d) $\SI{330}{\kelvin}$.}
 
  \label{fig:fig5}
\end{figure}

To determine an exact parameterization at each considered temperature, we conduct $\SI{250}{\nano\second}$ simulation run of $30$ monomer units long isotactic PNIPA. We use the radius of gyration as the target parameter; its lower value means that PNIPA is in the globule state, while a higher radius indicates the coil state. The results for either TC$2$-TP$5$-TC$5$, TC$2$-TP$4$-TC$5$ or TC$2$-TP$5$-TC$4$ imply that PNIPA is in the coil state within the entire temperature range under investigation, while TC$2$-TP$3$-TC$5$ remain in the globule state. In such case, we have to choose TC$2$-TP$5$-TC$5$ representation at temperatures below LCST and TC$2$-TP$3$-TC$5$ at temperatures above LCST.

\section{Discussion}\label{sec:D}

The main finding of this work is another confirmation that Martini force-field is just as powerful for modeling molecular machines as classic molecular dynamics. Although it might have some limitations, its overall credibility is strong as it is able to capture such effects as thermoresponsibility, spontaneous vibrations and stochastic resonance. 

Our new PNIPA parametrization in general maintains the characteristics of the previous one. However, since the Lennard-Jones parameters for the beads are different between the beta and release versions of Martini, some modifications are to be made. The main difference is that for the current model we use TC$2$-TP$5$-TC$5$ representation at all temperatures below LCST and TC$2$-TP$3$-TC$5$ at all temperatures above LCST (in our previous work we had to use TC$2$-TP$4$-TC$5$ representation at temperatures between $\SI{280}{\kelvin}$ and $\SI{300}{\kelvin}$). Note that although beads' denomination is mostly the same, the exact Lennard-Jones parameters are different.

Still, our results suggest that different representations must be used at temperatures below and above of LCST of PNIPA. Although we do not consider this a major flaw of the current model, we do think that the model might be further fine-tuned. First of all, we draw attention to the concept of A-mapping that we exploit for our model. This concept was absolutely necessary at the time when Martini $2.0$ was widespread since that version did not include tiny beads; small beads had to be used for both $3$-to-$1$ and $2$-to-$1$ mappings. In this case, A-mapping and B-mapping were used to distinguish nonmethylated polymers, such as polystyrene, and methylated polymers, such as polymethyl methacrylate\cite{rossi_coarse-graining_2011,uttarwar_study_2013}. With the introduction of tiny beads, the benefits of using these mappings are questionable. On the other hand, we cannot help but notice that switching to B-mapping implies reparameterization of intramolecular potentials, such as bond-length and bond-angle; moreover it also requires an introduction of dihedral angles. Thus, we preserve the A-mapping for our model, although we plan to investigate the B-mapped model in future.

Another possible fine-tuning of our model is the introduction of partial charges, which is allowed in the latest version of Martini. The results of \citeauthor{kamath_thermodynamic_2013} show that both the amide and isopropyl groups of PNIPA carry a partial charge, and the reflection of such a fact in the model might significantly improve it. However, we prefer to abstain from that, since once again it would imply reparameterization of intramolecular potentials.

The results obtained from the oligo-PF-$5$ foldamer verify its bistable nature even in a hydrophobic solvent. Once again, as shown in our previous work\cite{muratov_coarse-graining_2024}, spontaneous vibrations occur with pulling forces smaller than those in classical molecular dynamics\cite{astakhov_spontaneous_2024}. Still, we do not consider this to be a significant drawback of the CG model since all the other characteristics of bistability remain the same.

Our results suggest that larger systems, for example, several oligo-PF-$5$ foldamers coupled by a graphene plate dissolved in tetrahydrofuran\cite{frolkina_collective_2025}, may also be explored using Martini $3.0$. Atomistic simulations of such systems, as shown by theoretical calculations, may be incapable of capturing the necessary timescales, whereas coarse-grained ones seem to be able to cover the needed times. In this regard, the ability of the Martini $3$ force field to reproduce bistable effects in both hydrophilic and hydrophobic solvents appears to be the major advantage for the investigation of such systems. 
 
\section{Conclusion}\label{sec:C}

We have performed coarse-grained modeling of thermosensitive compound, PNIPA, in water, and of helix-like compound, oligo-PF-$5$, in tetrahydrofuran. Both results correspond well with atomistic modeling. For oligo-PF-$5$, we have investigated the conditions in which bistable effects, spontaneous vibrations and stochastic resonance, occur. These findings validate the latest Martini model as a promising tool for modeling of molecular machines and offer new possibilities in simulations of large constructions.
\section*{Acknowledgments}
 Authors thank Alexey Astakhov, Vladimir Bochenkov, Maria Frolkina, Vladislav Petrovskii, Anastasia Markina, and Alexander Valov for helpful discussions.

\bibliographystyle{unsrtnat}  
\bibliography{martini}

\end{document}